\documentclass[floatfix,nofootinbib,twocolumn,showpacs,superscriptaddress, reprintnumbers,amssymb,letterpaper,amsmath,pra]{revtex4-2}

\usepackage{graphicx}
\usepackage{dcolumn}
\usepackage{bm}
\usepackage{xcolor}
\usepackage[colorlinks=true,allcolors=blue]{hyperref}
\usepackage{amsmath}
\usepackage{lineno}
\usepackage{float}
\usepackage{tablefootnote}

\newcommand{\mean}[1]{\mbox{$\langle{#1}\rangle$}}
\usepackage{array}
\newcolumntype{C}[1]{>{\centering\arraybackslash}p{#1}}
\newcommand\hess{{\cal S}}

\begin{document}


\title{Numerical Modeling of a Proof-of-Principle Experiment on \\
Optical Stochastic Cooling at the IOTA Electron Storage Ring}
\author{A. J. Dick}
\email{adick1@niu.edu}
\affiliation{Northern Illinois Center for Accelerator \& Detector Development and Department of Physics, Northern Illinois University, DeKalb, IL 60115, USA} 
\author{M. Borland}
\affiliation{Argonne National Laboratory, Lemont, IL 60439, USA}
\author{J. Jarvis}
\affiliation{Fermi National Accelerator Laboratory, Batavia, IL 60510, USA}
\author{V. Lebedev}
\affiliation{Fermi National Accelerator Laboratory, Batavia, IL 60510, USA}
\author{P. Piot}
\affiliation{Northern Illinois Center for Accelerator \& Detector Development and Department of Physics, Northern Illinois University, DeKalb, IL 60115, USA} 
\affiliation{Argonne National Laboratory, Lemont, IL 60439, USA}
\author{A. Romanov}
\affiliation{Fermi National Accelerator Laboratory, Batavia, IL 60510, USA}
\author{M. Wallbank}
\affiliation{Fermi National Accelerator Laboratory, Batavia, IL 60510, USA}

\date{\today}

\begin{abstract}
Cooling of beams circulating in storage rings is critical for many applications including particle colliders and synchrotron light sources. A method enabling unprecedented beam-cooling rates, optical stochastic cooling (OSC), was recently demonstrated in the IOTA electron storage ring at Fermilab [\href{https://doi.org/10.1038/s41586-022-04969-7}{J. Jarvis, et al., Nature 608, 287–292 (2022)}]. This paper describes the numerical implementation of the OSC process in the particle-tracking program {\sc elegant} and discusses the validation of the developed model with available experimental data. The model is also  employed to highlight some features associated with different modes of operation of OSC. The developed simulation tool should be valuable in guiding future configurations of optical stochastic cooling and, more broadly, modeling self-field-based beam manipulations.
\end{abstract}

\maketitle


\section{introduction}
Particle-beam cooling~\cite{sessler-2006-a,poth-1990-a} $-$ the process of reducing beam emittances $-$ has been an essential element in the success of accelerator-based science~\cite{campagnari-1997-a,didella-2015-a}. For example, in the case of particle colliders, beam cooling is critical for increasing and preserving the collider's luminosity. A broad range of cooling techniques have been developed~\cite{poth-1990-a,budker-1978-a,derbenev-2000-a,aspect-1988-a,blaskiewicz-2014-a}. Of particular importance, the method of stochastic cooling (SC)~\cite{mohl-1980-a,bisognano-1983-a} enabled the production of intense antiproton beams, which subsequently led to the discovery of the $W$ and $Z$ bosons in 1983 at the Super Proton Synchrotron (SPS)~\cite{vandermeer-1985-a}. SC uses an electromagnetic pickup to measure position information associated with short longitudinal slices of the bunch~\cite{vandermeer-1987-a}. This information is encoded in the temporal structure of a radio frequency (RF) signal, typically in the microwave regime, amplified, and then used to correct the slices' average positions at a downstream kicker device. The correction is applied turn-by-turn as the beam circulates in the ring. At each turn, the sampled slices contain different randomized groupings of particles. Over many turns, each particle's coherent contribution to its own cooling signal comes to dominate over the diffusive contribution of neighboring particles, and thus, the beam is ``stochastically" cooled~\cite{mohl-1988-a}.

SC has been successfully implemented at many accelerator facilities for both particle accumulation and cooling~\cite{vandermeer-1972-a,vandermeer-1985-a,vandermeer-1985-b,mohl-1988-a,marriner-2004-a}. SC has also been successfully demonstrated and operationalized at the collision energy in the Relativistic Heavy Ion Collider (RHIC) for moderate-intensity bunched beams of both protons and gold ions~\cite{blaskiewicz-2007-a,blaskiewicz-2010-a}. Unfortunately, due to its limited bandwidth, microwave SC becomes ineffective at the higher intensities that are typical of, e.g., proton-(anti)proton colliders.

In the case of optimal cooling, where the system gain is balanced against diffusive effects, the maximum damping rate is approximately
\begin{eqnarray}
\tau^{-1} \simeq \frac{\Delta f \sigma_s}{NC},
\end{eqnarray}
where $\Delta f$ is the bandwidth of the integrated system, $N$ the number of particles in the bunch, $\sigma_s$ the root-mean-square (RMS) bunch length, and $C$ is the circular-accelerator circumference. For reference, with a typical proton beam in the Large Hadron Collider ($\sigma_s\simeq 10$~cm, $C\simeq 30$~km, $N\simeq 10^{11}$), an SC system with a bandwidth $\Delta f \simeq 4\times10^{9}$~Hz, would have a damping time on the order of $\tau \simeq 2\times10^3$~hours. \\

Effective cooling of such beams would require an increase in damping rates by at least three orders of magnitude. A possible solution is to significantly increase the bandwidth of the integrated system. One such extension of SC to optical frequencies and bandwidths, optical stochastic cooling (OSC), was proposed in the early 1990s~\cite{mikhailichenko-1993-a}. The OSC mechanism can support an optical bandwidth of up to $\Delta f \sim 100$~THz, potentially increasing the achievable damping rate by four orders of magnitude. The basic principles of OSC are similar to those of SC; however, OSC replaces the conventional microwave pickups and kickers with undulator magnets that deflect the beam particles to produce and couple to optical radiation. This radiation can be amplified using an optical amplifier similar to those employed in high-power free-space lasers. The possible use of OSC to cool hadron~\cite{lebedev-2014-a,babzien-2004-a} and lepton~\cite{zholents-2001-a,franklin-2007-a,andorf-2020-a,zholents-2021-a} beams has been proposed, and OSC-based techniques capable of manipulating the beam, e.g. halo control, have also been explored~\cite{zholents-2000-a,dick-2022-a}.

The OSC technique was first demonstrated in 2021 at Fermilab's IOTA storage ring using low-charge electron beams~\cite{jarvis-2022-a}. The original experiment implemented a ``passive" version of OSC where optical radiation emitted in the pickup interacts with the beam in the kicker without prior amplification. An amplified version of the experiment is currently under development at Fermilab.

To guide OSC R\&D activities, we have developed a computational model of OSC that can simulate the full OSC mechanism on a turn-by-turn basis. The model was benchmarked against the data collected during the passive OSC experiment at the IOTA ring. This model supplements the previous analytical models of OSC and will aid in the study of various aspects of the phase-space dynamics that are inaccessible in experiments. It will also serve as a valuable tool in the development of techniques for advanced beam control.
%
%
\section{Transit-Time OSC (TTOSC)}

\subsection{Theoretical Background~\label{sec:theory}}
\subsubsection{General principle}
OSC extends the principle of SC to optical frequencies by using undulator magnets for the pickup and kicker~\cite{mikhailichenko-1993-a,zholents-2012-a}. For a planar undulator, the on-axis fundamental wavelength of undulator radiation (UR) is given by
\begin{eqnarray}
\lambda = \frac{\lambda_u}{2\gamma^2} \left(1 + \frac{K^2}{2}\right)
\end{eqnarray}
where $\lambda_u$ is the undulator period, $\gamma$ is the Lorentz factor, and $K\equiv \frac{e\lambda_u B_u}{2\pi mc}$ is the undulator parameter where $e$ and $m$ are respectively the elementary charge and the particle's rest mass, $c$ the speed of light, and $B_u$ the peak magnetic field in the undulator~\cite{kim-1989-a}.

The original OSC concept included a quadupolar pickup undulator to produce emission based on the fluctuations of each sample slice, a form similar to conventional SC~\cite{mikhailichenko-1993-a}. In subsequent work, it was shown that OSC could be implemented using conventional undulators separated by a delay line that provides an energy-dependent path length for the particles -- a configuration referred to as transit-time OSC (TTOSC)~\cite{zolotorev-1994-a}.
\begin{figure}[h!]
    \centering
    \includegraphics[width=\columnwidth]{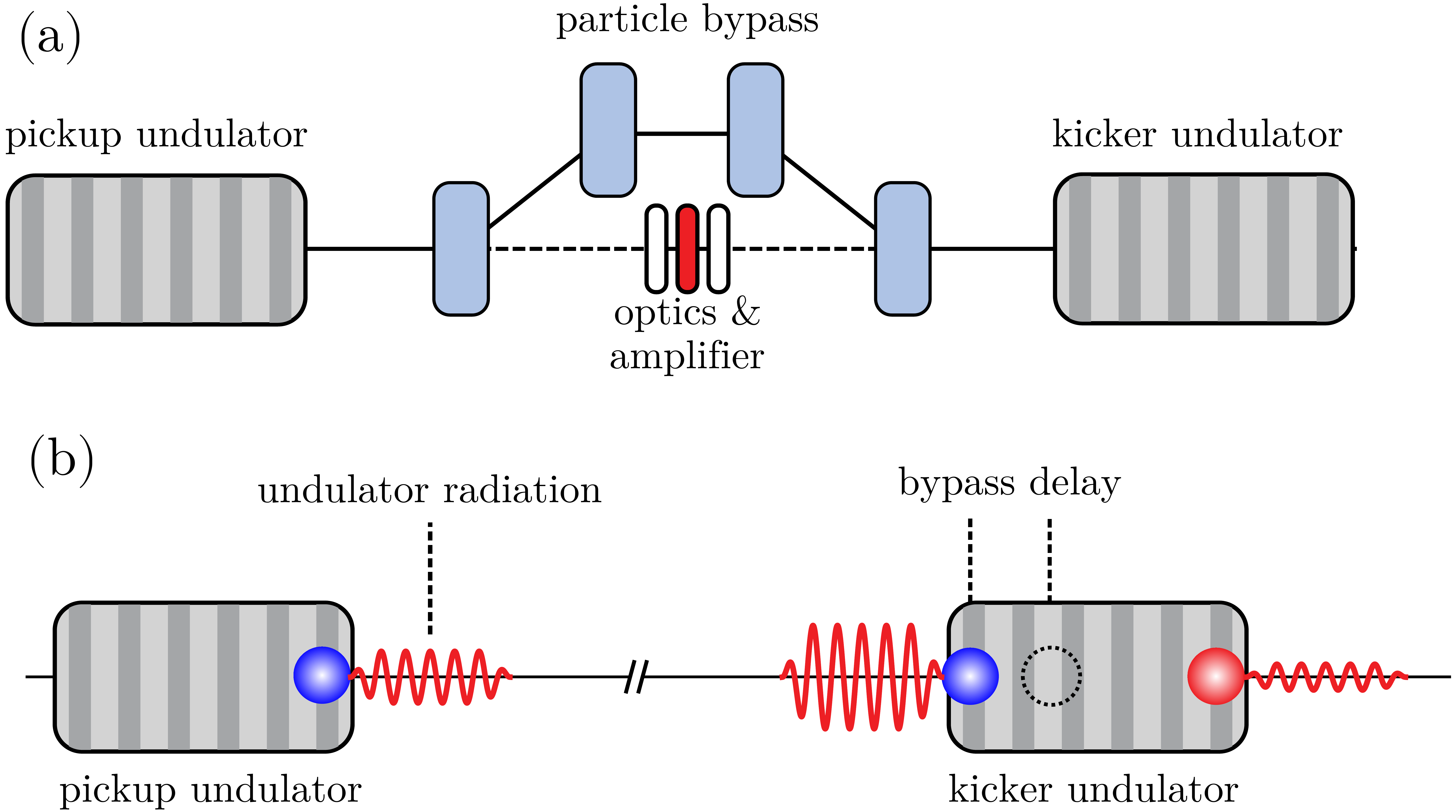}
    \caption{(a) A TTOSC system with pickup and kicker undulators, a bypass chicane, an optical line, and (b) the position of a particle and its own undulator radiation at the exit of the pickup undulator, and at the entrance and exit of the kicker undulator. The particle slips behind the radiation one wavelength each undulator period and the bypass introduces a delay so that the particle arrives at the kicker at the front of the radiation wavepacket. The delay can be tuned such that the subsequent energy exchange in the kicker produces corrective kicks that cool the beam.}
    \label{fig:osc_schematic}
\end{figure}
Figure~\ref{fig:osc_schematic} depicts the TTOSC concept: the pickup (PU) and kicker (KU) undulators are separated by a bypass beamline which introduces a nominal pathlength $s_0$ for the reference particle. A particle with phase-space coordinate $\mathbf{\cal Z}_{PU,i} \equiv (x_{PU,i}, x'_{PU,i}, y_{PU,i}, y'_{PU,i}, s_{PU,i}, \delta_{PU,i})^T$ at the PU location (with respect to the reference particle) will experience a modified path length~\cite{zolotorev-1994-a,ng-2003-a,lee-2004-a}
\begin{eqnarray}\label{eq:pathlengthNew1}
s_i &=&s_0+R_{51}x_{PU,i}+R_{52}x'_{PU,i}+R_{56}\delta_{PU,i}, \\
&=& s_0 + \Delta s(\mathbf{\cal Z}_{PU,i})
\end{eqnarray}
where $R_{5j}$ refers to the elements of the 6x6 transfer matrix between the PU and KU, $(x_{PU,i},x'_{PU,i})$ are the horizontal phase-space coordinates and $\delta_{PU,i}$ is the fractional momentum deviation. Equation~\ref{eq:pathlengthNew1} considers a bypass beamline acting on the 4D phase space $(x, x', s, \delta)$ so that the motion in $(y,y')$ is decoupled. We introduce the phase shift associated with the $i$-th particle as
\begin{eqnarray}\label{eq:pathlengthNew}
\Delta\phi_i= k (s_i-s_0)=\omega (\Delta t_i-\Delta t_0),
\end{eqnarray}
where $k\equiv \frac{2\pi}{\lambda}=\frac{\omega}{c}$ and $\Delta t_i$ and $\Delta t_0$ are the time of flight for the $i$-th particle and the reference particle respectively (i.e. $\Delta t_i=\frac{s_i}{\beta c}$ where $\beta\equiv \sqrt{1-\gamma^{-2}}$).

The radiation emitted by the particle in the PU propagates to the KU through an optical transport line which includes the optical amplifier, imaging lenses, and a delay-control system~\cite{andorf-2018-c}. The optical line introduces an optical path length (OPL) $\ell$. The delay-control system must maintain an OPL so that the optical delay $\delta \ell \equiv s_0 - \ell = \lambda/4$. This ensures that the reference particle's periodic maxima in transverse velocity are in phase with the zero crossings of the PU-radiation field, thus resulting in zero net energy exchange.

The energy change experienced by a particle in a single pass will depend on its phase difference with respect to the reference particle $\Delta \phi_i$ as~\cite{andorf-2018-c}
\begin{eqnarray}
    \Delta {\cal E }_i &=& e \int_0^{t_u}  \mathbf{E} \cdot \mathbf{v}  dt =  \frac{e}{c}\int_0^{L_u}  E_{x} v_x ds \nonumber \\
     &\simeq & - {\cal K} \sin(\Delta \phi_i + \psi_0),
    \label{eq:tt_coh}
\end{eqnarray}
where ${\cal K}$ represents the maximum possible energy change, $L_u$ is the length of the undulators, and $\psi_0$ is the phase offset which represents a temporal shift between the beam and radiation packet and is described further in Section~\ref{sec:modeling_osc}. The strength of the focused PU-radiation field is computed at the KU location. The latter equation also assumes that $s_i-\ell \le \lambda$. Equation~\ref{eq:tt_coh} represents the interaction of a particle with its own PU radiation while passing through the KU; this is commonly referred to as the coherent contribution to the kick. The coherent kick is the one that ultimately contributes to phase-space cooling. The UR wave packet produced by a particle as it passes through the PU can be approximated as a truncated sinusoidal wave with a total duration $N_u\lambda/c$. By the end of the PU the wave packet is located just ahead of the particle; see Fig.~\ref{fig:osc_schematic}(b). This slippage length $\hess\equiv N_u\lambda$  arises because the radiation propagates faster than the particle, which ``slips” by one optical wavelength ($\lambda$) per undulator period~\cite{pellegrini-2016-a}. In order to maximize the overlap of the interaction within the KU, the particle needs to arrive just ahead of the wave packet; see Fig.~\ref{fig:osc_schematic}(b). If the unwrapped phase is larger than $2\pi$ (i.e. $s_i-\ell > \lambda$) then the kick strength ${\cal K}$ reduces as discussed in Section~\ref{sec:coherentkick_algo}. \\


The cooling mechanism in TTOSC relies on the particles' longitudinal dynamics between the PU and KU. By its nature, TTOSC only cools the longitudinal phase space; however, the cooling force can be redistributed to the other phase planes by modifying the bypass and lattice optics. Generally, the cooling process leads to an emittance reduction parameterized as
\begin{eqnarray}
\varepsilon_m (t) = \left( \varepsilon_{m,0}-\varepsilon_{m,\infty} \right) e^{-\frac{t}{\tau_m}}+ \varepsilon_{m,\infty},
\label{eq:emit_diffeq}
\end{eqnarray}
where $m=x,y,s$ refers to the considered degree of freedom, $\varepsilon_{m,0}$ and $\varepsilon_{m,\infty}$ are respectively the injection and equilibrium emittances, and $\tau_m$ is the damping time. Throughout this paper, the emittance values correspond to the geometric emittances statistically computed from the macroparticle distribution~\cite{lapostolle-1971-a} as 
\begin{eqnarray} \label{eq:emit}
\varepsilon_m\equiv [\mean{m^2}\mean{m'^2}-\mean{mm'}^2]^{1/2}.
\end{eqnarray}
Generally, $\tau_m={\cal J}_m^{-1} \tau_0$ with the partition numbers ${\cal J}_m$ connected by ${\cal J}_x+{\cal J}_y+{\cal J}_s=4$~\cite{robinson-1958-a} and $\tau_0$ being the damping time associated with the considered energy-loss process (e.g. synchrotron radiation and/or OSC in the present work). Consequently, the damping decrements satisfy
\begin{eqnarray}
\tau_x^{-1}+\tau_y^{-1}+\tau_s^{-1}= 4 \tau_0^{-1}\equiv \tau^{-1}_{tot}.  
\end{eqnarray}
The ratio between damping decrements associated with the different phase-space planes can be controlled by the horizontal dispersion and the lattice coupling. Specifically, the cooling force can be shared with the horizontal plane by introducing dispersion in the undulators and modifying the PU-KU beamline settings to control the path-length dependence on horizontal coordinates $(x,x')$ in Eq.~\ref{eq:pathlengthNew1}. The cooling force can be shared with the vertical plane by coupling the horizontal and vertical planes using skew-quadrupole magnets or by operating the storage ring on a transverse-coupling resonance.

\subsubsection{Multi-particle interactions}
In addition to the coherent kick discussed above, a particle also experiences the field produced by other particles whose radiation wave packets overlap with the considered particle. These ``incoherent" contributions of the neighboring particles can be derived similarly with an additional phase difference due to their relative separation.

The radiation wave packet produced by the $j$-th particle interacts with the $i$-th particle in the KU if $\left|t_i-t_j\right| < N_u\lambda/c$, where $t_i$ and $t_j$ are the particles' arrival times at the PU center. Introducing the relative phase between the particles, $\phi_{ij} = \omega(t_i - t_j)$, the effective kick produced by the $j$-th particle on the $i$-th particle can be calculated in the same way as Eq.~\ref{eq:tt_coh} using $\phi_{ij}$ as the relative phase term. The total incoherent kick experienced by the $i$-th particle is then obtained by summing over all particles having time coordinates within the temporal slice, i.e. $t_j \in [ t_i- \hess/c,  t_i+\hess/c]$, and is given by
\begin{eqnarray}\label{eq:ttosc_incoherent}
    \Delta \widetilde{\cal E}_{j} = -{\cal K}  \sum_{j\neq i}\sin(\Delta \phi_i + \phi_{ij} + \psi_0).
\end{eqnarray}
The incoherent kick  for a given particle thus depends on the number of particles within a slippage length to either side and corresponds to the summation of the wave packets produced by each particle~\cite{kim-1997-a}.

\subsubsection{Transverse effects\label{sec:singleparticleTransEffect}}

So far our description of the TTOSC method has assumed the UR wavepacket to be a plane wave. We may also account for the non-uniform transverse field distribution of UR  and the particles' transverse motion between the PU and KU. The UR is emitted within a cone with apex angle $\theta\sim 1/\gamma$ and is then imaged in the kicker. Ignoring prefactors and assuming an undulator parameter $K\ll 1$~\cite{andorf-2018-b,andorf-2018-c}, the far-field distribution of the electric field in the KU has the form
\begin{equation}\label{eq:trans_field}
\begin{split}
    e_x (r, \phi) \propto \frac{\theta \left[J_0\left(\xi\right) + (\gamma\theta)^2\cos(2\phi) J_2\left(\xi\right)\right]}{(1+(\gamma\theta)^2)^4},
    \end{split}
\end{equation}
where $\xi\equiv \frac{r k  \theta}{1+(\gamma\theta)^2}$, $J_0$ and $J_2$ are Bessel functions of the first kind of order $n=0,2$ respectively, and $\theta$ is the polar angle of a given ring-shaped surface element on the focusing lens relative to the center of the PU.
The horizontal component of the electric field at the focal point in the KU is then proportional to the integral over the angular acceptance $\theta_m$ of the lens,
\begin{eqnarray}
    E_x(x,y) \propto \int_0^{\theta_m} e_x(r,\theta) d\theta,
    \label{eq:off_axis_field}
\end{eqnarray}
where $r \equiv \sqrt{x^2+y^2}$. In the absence of depth-of-field effects, the UR from the PU is imaged into the KU with a total transfer matrix $M=-I$, where $I$ is the identity matrix~\cite{nazarathy-1982-a}. If the transfer matrices for the particles and light are not matched in the transverse planes, then the particles may arrive in the KU off-axis relative to their PU radiation. The relative transverse offset of a particle is determined by the difference in transverse coordinates at the pickup and kicker. The offset of a single particle is defined as $\Delta x_i=x_{U,i} - M x_{PU,i}$. The electric field sampled by the $i$-th particle in the PU is then $E_x(\Delta x_i, \Delta y_i)$. The correction for the total energy change experienced by an off-axis particle is given by substituting ${\cal K} \rightarrow {\cal K} \varrho(x,y) $ in Eq.~\ref{eq:tt_coh} where  
\begin{eqnarray} \label{eq:off_axis_rho}
\varrho(x,y)=E_x(x,y)/E_x(0,0).
\end{eqnarray}

It should be noted that, for a given OSC system, the finite angular acceptance of the optical system also decreases the maximum energy change~\cite{andorf-2018-b}
\begin{eqnarray}
   \mathcal{K}(\theta_m) = \frac{\pi}{3\epsilon_0\lambda_0} e^2 N_u F_T(K,\gamma\theta_m),    
\end{eqnarray}
where the suppression factor $F_T()$ depends on the angular acceptance of the optical system $\theta_m$ as detailed in~\cite{andorf-2018-c}. Therefore the transverse effects leads to Eq.~\ref{eq:tt_coh} being modified as
\begin{eqnarray}
    \Delta {\cal E}_i (x,y)=-\varrho(x,y) \mathcal{K}(\theta_m) \sin(\Delta \phi_i + \psi_0).
\end{eqnarray}
The previous equation together with Eq.~\ref{eq:ttosc_incoherent} assumes a passive-cooling configuration. When an optical amplifier is included both of these equations should include a multiplicative factor $\sqrt{G}$, where $G$ represents the optical-power gain.
\subsection{OSC proof-of-principle experiment configuration at IOTA}
\begin{figure} [hhhhh!!!!!] 
    \centering\includegraphics[width=\columnwidth]{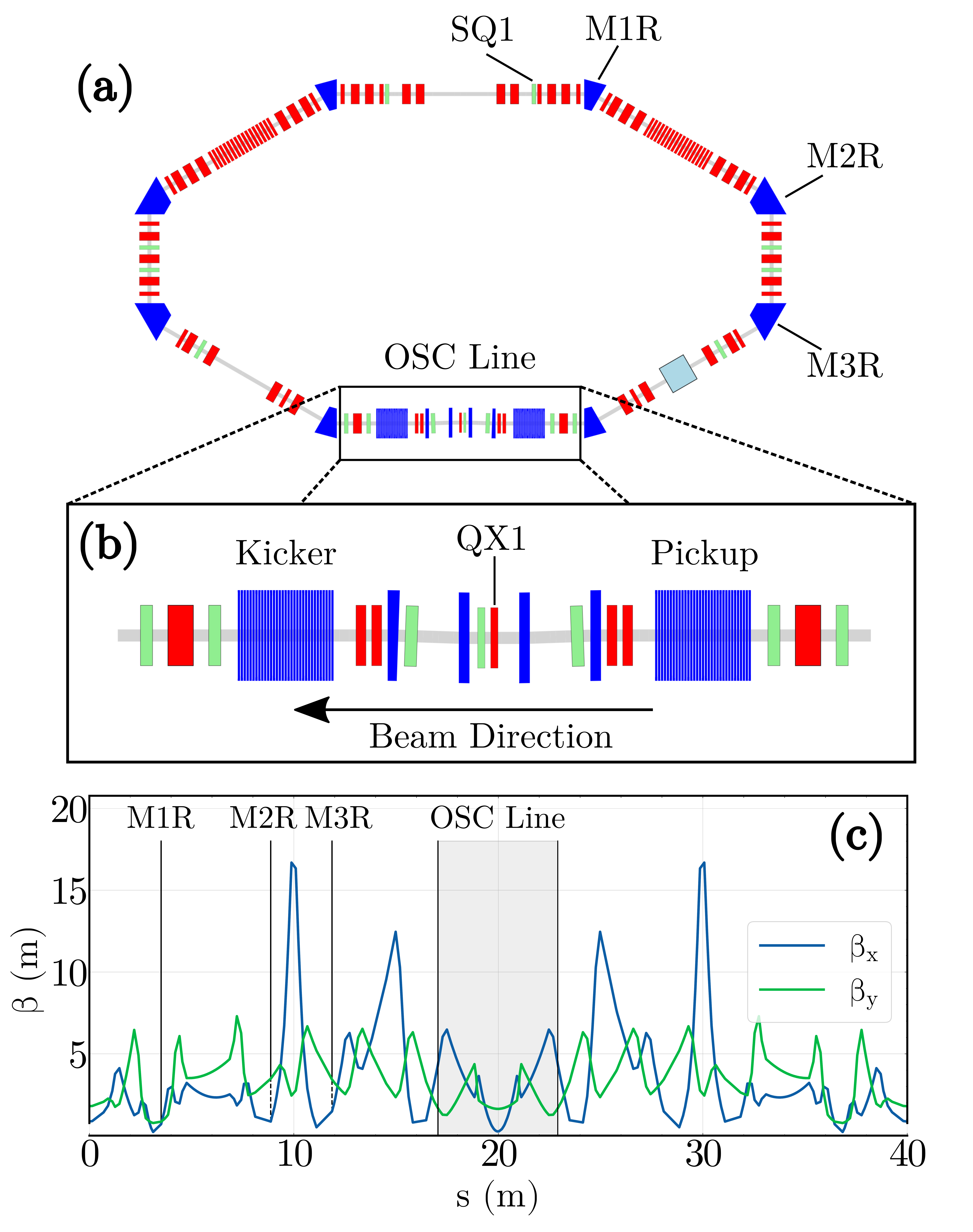}   
   \caption{Diagram of the IOTA storage ring at Fermilab (a) with a zoomed-in view of the OSC line (b) and horizontal and vertical betatron functions around the ring (c). In (a,b) blue, red, and greed shapes correspond respectively to dipole, quadrupole, and sextupole magnets. In (c), the shaded area corresponds to the location of the OSC section and the labels M$i$R ($i=1,2,3$) indicate the location of the first three bending magnets [see (a)] where beam diagnostics are available. The beam circulates in the clockwise direction in the ring (i.e. SQ1$\rightarrow$ M1R $\rightarrow$ M2R $\rightarrow$ M3R$\rightarrow$ OSC $\rightarrow$ SQ1). }
    \label{fig:iota_diagram}
\end{figure}

The OSC process was experimentally demonstrated at Fermilab's IOTA storage ring~\cite{antipov-2017-a,lebedev-2014-b,lebedev-2021-a} and data collected during the experiment~\cite{jarvis-2022-a} are used to benchmark the computational model in the next sections. Table~\ref{tab:OSCPOP} summarizes the main experimental parameters and a diagram of the system appears in Fig.~\ref{fig:iota_diagram}(a). The OSC bypass includes four dipole magnets arranged as a chicane, four quadrupole magnets for control of beam focusing and dispersion, a single quadrupole (QX1) to control the coupling of OSC between the longitudinal and horizontal planes, and three sextupole magnets for mitigation of nonlinear path lengthening in the bypass. 
\begin{table}[hhhhh!!!]
    \caption{Nominal undulator and beam parameters for the passive TTOSC proof-of-principle experiment at the IOTA facility.\label{tab:OSCPOP}}
    \vspace{3mm}
\begin{tabular}{l l c}
\hline
\hline
parameter, symbol & value & unit. \\
\hline
undulator parameter, $K$         & 1.038 & -\\
length,  $L_u$    & 77.4 & cm           \\
undulator period,  $\lambda_u$    & 4.84 & cm \\
number of periods, $N_u$    & 16 & - \\
on-axis wavelength, $\lambda$ & 950 & nm \\
total bypass delay  & 0.648 & mm \\
\hline
electron beam energy, $E_b$ &    100  & MeV \\
electron beam current, $I_b$ & [50, 150] & nA \\ 
electron Lorentz factor, $\gamma$    & 196.69 & - \\
maximum energy change, ${\cal K}$ &   60 & meV \\ 
\hline
\hline
\end{tabular}
\end{table}

For the uncoupled lattice [with computed lattice functions displayed in Fig.~\ref{fig:osc_schematic}(c)], OSC only cools the longitudinal phase space; however, the cooling force can be distributed between the longitudinal and horizontal planes by powering QX1, which is located at the symmetry point of the particle bypass with significant horizontal dispersion (see Fig.~\ref{fig:iota_diagram}). Additionally, the cooling force can further be shared with the vertical plane using the skew-quadrupole magnet SQ1 that introduces coupling between the $(x,x')$ and $(y,y')$ transverse phase spaces~\cite{lebedev-2021-a}. This latter configuration supports three-dimensional cooling. These two quadrupole magnets were the only lattice elements tuned during the benchmarking process to investigate the cooling dynamics in the three coupling configurations.

The primary metrics on which we evaluated the model were ($i$) the cooling rates of various lattice configurations and ($ii$) the equilibrium longitudinal and transverse beam distributions. A complete lattice of the IOTA ring was simulated in \textsc{elegant}~\cite{borland-2000-a} using magnet settings from the experiment lattice; see Fig.~\ref{fig:iota_diagram}(c).


\section{Modeling OSC in Elegant\label{sec:modeling_osc}}
We opted to implement the OSC process in the {\sc elegant} particle-tracking program owing to its large user base and open-source nature~\cite{borland-2000-a}.  {\sc Elegant} has been extensively used for the design of storage rings and linear accelerators. Particle tracking in {\sc elegant} can be performed using matrices of selectable order, canonical kick elements, numerically integrated elements, or any combination thereof. Canonical kick elements are available for dipole, quadrupole, and sextupole magnets, and higher-order multipoles. All of these elements support optional classical synchrotron radiation losses, as well as random variation in radiation, leading to quantum excitation. The particle-tracking implementation of most of the beamline elements is parallelized on CPU~\cite{wang-2006-a} and some have GPU capabilities~\cite{pogorelov-2014-a}. The code also supports parallel simulation of multi-bunch beams with short- and long-range wakefields~\cite{borl:icap15-webji1}, beam-loaded rf cavities with feedback~\cite{berenc:ipac15-mopma006}, and bunch-by-bunch transverse and longitudinal feedback, all of which are of interest in modeling realistic beam behavior. The parallelization of {\sc elegant} is critical to performing detailed simulations of the long-term beam dynamics and understanding the cooling process.

\subsection{General considerations}
The main components of the OSC implementation in {\sc elegant} are the {\tt cpickup} and the {\tt ckicker} beamline elements. Both are zero-length elements. The {\tt cpickup} records the internal 6-D coordinates of every macroparticle while the {\tt ckicker} element applies a momentum kick. The two elements form a linked pair within {\sc elegant} using an identifier string. In its simple implementation, OSC is modeled by placing a {\tt cpickup} and  {\tt ckicker} in the center of respectively the KU and PU. Such a ``single" kick approximation captures most of the features associated with the OSC mechanism. However, the kick can also be applied in a distributed fashion using multiple {\tt cpickup}-{\tt ckicker} pairs distributed along each of the undulators. Such a configuration allows for the simulation of angular misalignment, and for investigating the impact of optical focusing and magnification errors. {\sc Elegant} uses the 6-D coordinates $(x,x',y,y',s,\delta)$ where $s$ is the total, equivalent distance traveled, and $\delta\equiv (p-p_0)/p_0$ (where $p_0$ is the reference-particle momentum) is the fractional momentum deviation~\cite{brown-1985-a}. 

At present, the OSC model has three main components discussed in Section~\ref{sec:theory}: (i) coherent kicks, (ii) incoherent kicks, and (iii) transverse effects. The following three sections discuss the specifics of the implementation associated with these three processes.

\subsection{Coherent kicks \label{sec:coherentkick_algo}}
The coherent energy kick described by  Eq.~\ref{eq:tt_coh} is implemented as a relative momentum change experienced by the $i$-th macroparticle following
\begin{eqnarray}
    \frac{\Delta p_i}{p_0} &=& -\kappa \Psi(\Delta\phi_i+\psi_0)\sin(\Delta\phi_i+\psi_0)
    \label{eq:elegant_tt_coh}
\end{eqnarray}
where $\Delta\phi_i = \omega (\Delta t_i - \Delta t_0)$ and the variables $\kappa$, $\omega$, and $\psi_0$, are user-defined parameters for respectively the maximum normalized kick strength, radiation frequency, and unwrapped optical-delay phase. The function $\Psi()$ describes the temporal overlap as detailed below. The 6-D coordinates of each macroparticle are recorded in the PU using the {\tt pickup} element. At the location of the {\tt kicker} element, the time of flight from the location of the {\tt kicker} element is computed as $\Delta t_i=t_{i, kU}-t_{i,PU}$; this corresponds to the path length $s_i\simeq c \Delta t_i$ for ultra-relativistic macroparticles. The reference particle time of flight $\Delta t_0$ is computed either from the bunch average time of flight, $\Delta t_0 \simeq s_0/c=\mean{t_{KU}}-\mean{t_{PU}}$ where $\mean{...}$ indicates the statistical averaging over all the macroparticles, or from the closed orbit trajectory found by {\sc elegant}.

The delay phase, $\psi_0$, is an important parameter in the model as it affects both the phasing of the OSC effect and the temporal overlap of the beam and radiation pulse. Under normal operation, $\psi_0=0$ and a particle with zero momentum deviation will arrive in phase with its radiation, as illustrated in Fig.~\ref{fig:osc_schematic}(b); however, if there is an additional delay of $\psi_0=\pm\pi$ in the optical line, the reference particle arrives antiphased with its radiation and will receive a kick with the opposite sign and be driven away from the design momentum. As the UR is periodic, if the delay is an entire period ($\psi_0=\pm2\pi$) then the system will return to the cooling mode but with a slightly reduced cooling force due to the imperfect temporal overlap. The energy is only transferred from the radiation to the beam for as long as the two are in the kicker together. Thus, as the delay increases, the total energy transferred decreases. We model this effect using the unwrapped delay phase as
\begin{eqnarray}
\Psi(\phi) = \begin{cases}
      0 & |\phi| > 2\pi N_u  \\
      1-|\phi|/2\pi N_u & |\phi| \le 2\pi N_u
   \end{cases}
\end{eqnarray}
The nominal kick in Eq.~\ref{eq:tt_coh} is multiplied by $\Psi()$ yielding Eq.~\ref{eq:elegant_tt_coh}. This causes the kick strength to decrease linearly with the optical delay until it vanishes when the particle and its undulator-radiation wavepacket no longer overlap.

\subsection{Incoherent kicks}
As described above, the field experienced by a particle in the kicker is the superposition of the field produced by all particles in its sample slice $z_i \pm N_u\lambda$. In order to simulate the effect of neighboring particles, we consider their individual contributions to the corrective kick instead of their contributions to the electric field in the kicker~\cite{wang-2021-a}. Each neighboring particle will affect the $i$-th particle in a similar way to the coherent contributions but with an unwrapped phase term $\psi_{ij}$ determined by the difference in arrival time at the pickup.

The incoherent-kick contribution to the $i$-th particle is modeled by considering the contributions of all other particles $j$ within the slippage interval $|t_i - t_j| \le N_u\lambda/c$ at the PU.  The phase difference $\phi_{ij} = \omega(t_i - t_j)$ at the PU is combined with the coherent phase for particle $i$ at the KU to ensure the correct is kick received from each particle $j$, as described in Eq.~\ref{eq:ttosc_incoherent}.  The kick reduction due to the total optical delay described by $\Psi()$ also influences the incoherent kick.  In practice, these contributions become more significant as the longitudinal beam density increases but are negligible for the considered low-charge beams. In our numerical implementation in {\sc elegant}, this effect is parallelized and can be toggled in the lattice file using the {\tt incoherentMode} parameter.

\subsection{Transverse effects\label{sec:simu_transEffects}}
The implemented TTOSC model accounts for the effect of the transverse motion of particles on the time of flight described in  Eq.~\ref{eq:pathlengthNew}. Additionally, the macroparticle may enter the KU with a transverse offset thereby sampling off-axis regions of the electric field as discussed in Section~\ref{sec:singleparticleTransEffect}. We account for this effect by calculating the relative change of the field with respect to the on-axis field, $\varrho(x,y)$, sampled by the macroparticle with transverse offset $(x,y)$; see Eq.~\ref{eq:off_axis_rho}. The user-defined maximum kick parameter $\kappa$ in Eq.~\ref{eq:elegant_tt_coh} is modified as  
\begin{eqnarray}
\kappa \rightarrow \kappa(x,y)=\varrho(x,y)\kappa. 
\end{eqnarray} 

The transverse motion of particles is calculated by subtracting the position in the kicker from the position in the pickup. The transverse displacement between UR emitted from the {\tt pickup} and imaged in the {\tt kicker} element and the position of the $i$-th macroparticle in the {\tt kicker} element is given by 
\begin{eqnarray}
\begin{cases}
~\tilde{x}_i = x_{KU,i} - {\cal M} x_{PU,i} &  \\
~\tilde{y}_i = y_{KU,i} - {\cal M} y_{PU,i} & 
\end{cases}
\end{eqnarray}
where ${\cal M}$ is the optical magnification between the {\tt pickup} and {\tt kicker}. The magnification is a user-defined parameter (${\cal M}=-1$ by default, corresponding to a single-lens optical-imaging system such as employed in the IOTA experiment). The relative field is calculated for each macroparticle using Eq.~\ref{eq:off_axis_field} with the correction factor $\varrho(\tilde{x}_i,\tilde{y}_i)$. The model does not currently include the transverse effects for the incoherent contributions and instead uses the on-axis strength. This approximation is acceptable for modeling the IOTA experiment as the incoherent effects are negligible (owing to the low number of electrons per sampling slice). The algorithm for modeling the transverse effects is parallelized and may be toggled in the model using the {\tt  transverseMode} input parameter in {\sc elegant}.


\section{Comparison of Numerical and Experimental results\label{sec:benchmarking}} 
An important motivation for developing a precise numerical model of OSC is to gain insight into the single particle evolution usually inaccessible in experiments. {\sc Elegant} incorporates several diagnostics  that provide either ensemble-averaged information on the beam distribution or phase-space coordinates of each macroparticle at a given location on a turn-by-turn basis. In particular, {\sc elegant}'s {\tt watch} element was extensively employed to export macroparticle distributions at several locations around the IOTA storage ring. This capability permitted the reconstruction of the data collection methods used at IOTA while providing insight into the dynamics of individual macroparticles in the phase space.
\subsection{Measurements}
The IOTA experiment relies on several diagnostics systems to measure the beam distribution. The transverse $(x,y)$ distribution of the beam is captured by two complementary metal–oxide–semiconductor (CMOS) cameras that image the synchrotron radiation produced in the M1L and M2R dipole magnets; see Fig.~\ref{fig:iota_diagram}(a). The required exposure time depends on the beam intensity. Typical values were between 1 and 100 ms, corresponding to many thousands of beam revolutions. The temporal distribution is measured using a streak camera installed at the M3R diagnostics station. The streak camera uses a continuous sweep voltage and is phase-locked to the $11^{th}$ harmonic (82.5~MHz) of the circulation frequency. Correspondingly, the {\sc elegant} simulations are configured to record the macroparticle distributions at the center of the M3R bending magnet for direct comparisons with the measurement. The signal can be integrated over multiple turns to mimic the response time of the cameras.

{\sc Elegant} allows the user to toggle most advanced effects independently, e.g. synchrotron radiation energy loss and quantum excitation effects may be applied individually. This feature is also built into the model of OSC so that incoherent contributions and transverse effects can be added to the basic transit-time model. The measurements of emittance damping in Tables~\ref{tab:sr_rates}, \ref{tab:theory_rates}, and \ref{tab:exp_rates} do not include quantum excitation or other diffusive effects. This reduces random noise in the emittance calculation and does not impact the observed cooling rate as diffusive effects only increase the equilibrium term in Eq.~\ref{eq:emit_diffeq}.

\subsection{Comparison of the nominal IOTA lattice with {\sc elegant} simulations}
The lattice configuration and magnet strengths were taken from the IOTA experiment in the Summer of 2021.
The three coupling configurations can be controlled using the quadrupole magnet QX1 located in the OSC bypass chicane and the skew quadrupole magnet SQ1 near the dipole magnet M1R. The full ring and OSC beamline is diagrammed in Fig.~\ref{fig:iota_diagram}(a,b). The PU and KU are modeled as 16 periods of alternating dipole magnets with the two outermost dipoles on either side forming a standard [1/4,3/4] termination scheme~\cite{gottschalk-1996-a}; see Table~\ref{tab:OSCPOP}. \\

Prior to investigating the OSC process in detail, we compared results obtained from the {\sc elegant} model with available nominal values for some of the lattice parameters in the absence of OSC. The theoretical values were initially calculated for the designed lattice~\cite{lebedev-2021-a} and were recomputed, using various optics codes, for slight changes in the final lattice used in the experiment. The parameters of the final lattice are considered here and are summarized in  Table~\ref{tab:lattice_parameters}.

\begin{table}[hhh!!!]
    \caption{Comparison of lattice parameters between the nominal (``theory") values~\cite{lebedev-2021-a} and the simulated values obtained from the {\sc elegant} model. These calculations do not consider the OSC process.\label{tab:lattice_parameters}}
    \vspace{3mm}
    \centering
    \begin{tabular}{lccc}
         \hline
         \hline
         Parameter & units &  theory & simulation \\
         \hline
         Energy loss per turn &   eV          & 12.7       & 12.67\\
         Betatron tunes $(\nu_x/\nu_y)$ & $-$  & 5.42/2.42  & 5.42/2.42\textsuperscript{\textdagger} \\
         Chromaticity $(\xi_x/\xi_y)$  &  $-$       & -10.2/-8.1 & -13.2/-6.15\textsuperscript{\textdagger}\\
         Momentum compaction & nm          & 0.00493   &  0.00489\textsuperscript{\textdagger}\\
         Synchrotron frequency & Hz        & 426.5     &  428.1 \\
         \hline
         \hline
    \end{tabular}
    \\
    \textsuperscript{\textdagger} Computed in {\sc elegant} using lattice optics
\end{table}

During the experiment, the synchrotron-oscillation frequency was measured for RF-cavity voltages within $[70,140$]~V. It was ultimately set to 105~V corresponding to a synchrotron frequency of $f_s=426.5$~Hz in agreement with {\sc elegant} value within $3\%$. Likewise, the synchrotron-radiation damping dynamics and the equilibrium emittances were simulated with {\sc elegant}. This is particularly important for the OSC studies since the OSC damping rates and equilibrium emittances are directly compared to the SR damping ones. We simulate a beam initially above equilibrium and allow it to circulate for several seconds corresponding to $\sim 25 \times 10^6$ turns. The damping decrement and equilibria, fit from these tracking simulations, are in good agreement with theoretical estimates, indicating that the baseline lattice and performance are properly simulated in {\sc elegant}; see Table~\ref{tab:sr_rates}.
\begin{table}[hhhhhhhhhh!!!!]
    \caption{Comparison of synchrotron-radiation (SR) damping and equilibrium-distribution parameters analytically computed (``theory")~\cite{lebedev-2021-a} with simulated values obtained from the {\sc elegant} model (``{\sc elegant}") \label{tab:sr_rates}. These calculations do not consider the OSC process or lattice coupling. The emittance and bunch length values are RMS values.}
    \vspace{3mm}
    \centering
    \begin{tabular}{lcll}
         \hline
         \hline
         parameter (symbol) & units & theory & simulation \\
         \hline
         SR hori emit. damping rate ($\tau_x^{-1}$) & s$^{-1}$ & 0.944  & 0.90 \\
         SR vert. emit. damping rate ($\tau_y^{-1}$) & s$^{-1}$ &  0.986 &  0.95\\
         SR long. emit. damping rate ($\tau_s^{-1}$)   & s$^{-1}$ & 2.014       & 2.03 \\
         equilibrium hor. emit. ($ \varepsilon_{x,\infty}$)     & nm   & 0.778       & 0.87 \\
         equilibrium vert. emit. ($\varepsilon_{y,\infty}$)     & nm   & 0.0       &  0.0 \\
         equilibrium bunch-length ($\sigma_{s,\infty}$)  & cm   & 5.4          & 5.78 \\
         \hline
         \hline
    \end{tabular}
\end{table}

\subsection{Validation of the {\sc elegant} OSC model}

The simulations outlined here consider the theoretical and experimental cases separately. The simulations of the theoretical design use the nominal values for angular acceptance, kick strength, and coupling terms. In the latter simulations, we attempt to recreate the conditions of the experiment by reducing the angular acceptance of the focusing lens, the total kick strength of the OSC element, and the strength of the coupling quadrupole magnet. We also include a simple scattering element to mimic residual gas scattering.

\subsubsection{Comparison with design values}
In the theoretical design, the total emittance damping rate due to OSC, in the absence of synchrotron radiation, scattering, transverse field effects, and incoherent contributions, is $\tau^{-1}_{tot}\simeq 38$~s$^{-1}$ ~\cite{lebedev-2021-a}. These effects can each be disabled individually in the {\sc Elegant} model to give a clear picture of the unaltered OSC damping. When simulating the ``base'' OSC effects in this way, we measure a combined transverse damping rate of $\tau_{x,y}^{-1}=19.1~s^{-1}$ and a longitudinal damping rate of $\tau_s^{-1}=18.0~s^{-1}$ corresponding to a total rate in all planes of $\tau^{-1}_{tot} = 37.0~s^{-1}$.

Next we simulated OSC in each of the coupling configurations while including synchrotron radiation effects and the transverse field calculation. Beginning with the uncoupled lattice, we set the parameters of the pickup and kicker elements to their theoretical values. In this case, the total longitudinal damping rate increases to $\tau_s^{-1}=29.6 ~\mbox{s}^{-1}$ from the one initially computed with SR damping alone ($\tau_s^{-1}=2.03 ~\mbox{s}^{-1}$). The horizontal and vertical rates remain the same, as determined by SR damping. In the coupled cases, the nominal ratio of cooling rates between the transverse and longitudinal planes is $\tau^{-1}_s:\tau^{-1}_x = 1.00:1.03$. We then set the strength of QX1 to the theoretical value corresponding to 1:1.03 coupling. Here, the longitudinal damping rate decreases to $\tau_s^{-1}=16.35 ~\mbox{s}^{-1}$  and the horizontal damping rate becomes $\tau_x^{-1}=16.49~\mbox{s}^{-1}$. Finally, the SQ1 quadrupole is activated and the cooling force is seen in all three planes. The longitudinal damping rate is approximately the same, and the transverse cooling is now evenly split between the horizontal and vertical degrees of freedom with $\tau_x^{-1}=9.3~\mbox{s}^{-1}$ and $\tau_{y}^{-1}=9.2~\mbox{s}^{-1}.$ This demonstrates the expected redistribution of the total cooling force between the three phase-space planes as the coupling terms are introduced. Table~\ref{tab:theory_rates} summarizes the results of these simulations. The total damping rate, summed across all dimensions, is consistent between the three configurations. Subtracting the SR-damping rates in each case leaves the OSC rates, which are observed to be $\sim 25\%$ lower than the simulated total rate discussed above ($\tau^{-1}_{tot} = 37.0~s^{-1}$). This can be explained by the reduced kick strength that results from the transverse effects described in Section~\ref{sec:simu_transEffects}.

\begin{table}[hhh!!!]
    \caption{Total damping rates simulated with {\sc elegant} for the IOTA design configuration. \label{tab:theory_rates}}
    \vspace{3mm}
    \begin{ruledtabular}
    \begin{tabular}{lddd}
      &  \multicolumn{3}{c}{damping rates ($s^{-1}$)} \\
    \textrm{Mode} & \textrm{$\tau_x^{-1}$} & \textrm{$\tau_y^{-1}$} & \textrm{$\tau_s^{-1}$} \\
    \hline
        SR only & 0.9 & 0.95 & 2.03 \\
        \hline
        uncoupled (1D)& 0.90 & 0.95 & 29.57 \\
        s/x coupled (2D)& 16.49 & 0.95 & 16.35 \\
        s/x/y coupled (3D)& 9.30 & 9.20 & 15.33 \\
    \end{tabular}
    \end{ruledtabular}
\end{table}

\begin{figure}[hhhhh!!!!!]
    \centering
    \includegraphics[width=\columnwidth]{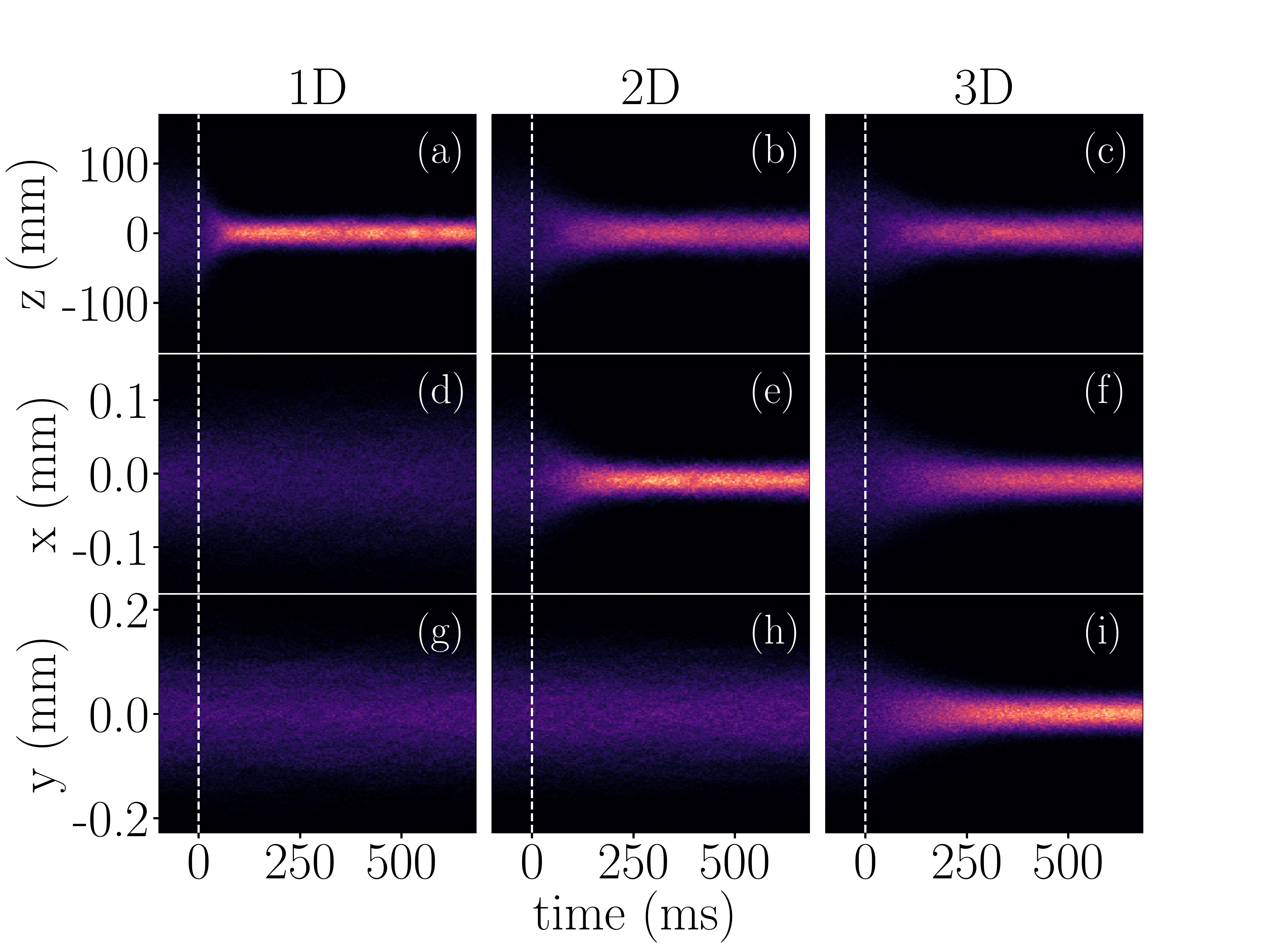}
    \caption{Waterfall plot of the temporal evolution of longitudinal (a) horizontal (b) and vertical beam distribution. OSC is turned on instantaneously at $t=0$~ms (indicated with the vertical dash lines). The ``1D", ``2D", and ``3D" columns correspond to lattice configured for respectively  longitudinal-only, longitudinal-horizontal, and three-dimensional cooling.}
    \label{fig:toggle_OSC}
\end{figure}

Figure~\ref{fig:toggle_OSC} presents the evolution of the projected distributions for the three coupling configurations. As expected, in the uncoupled case, only the longitudinal beam size is reduced. As the OSC force is shared with the other planes, the equilibrium size in the longitudinal plane grows larger while the equilibrium beam size in the transverse planes decreases thus confirming the redistribution of the  cooling process among the coupled planes.

\subsubsection{Comparison with IOTA experiment\label{sec:experimentsims}}
The IOTA experiment serves as an important benchmark for the computational model. In the experiment, the strength of OSC was characterized by measuring the projected beam distribution along each direction ($x$,$y$, and $z$) and accounting for the effects of intrabeam scattering (IBS) using a simple model \cite{jarvis-2022-a}. Relative to SR damping alone, the IOTA experiment measured an increase in damping of 8.06 times in the longitudinal and of 2.94 times in the transverse. This corresponds to a total OSC emittance cooling rate of $\tau_{tot}^{-1}=18.4$~s$^{-1}$ and is approximately half of the theoretical estimate. Several sources were identified as possible explanations for this difference \cite{jarvis-2022-a}. The first was a reduced, asymmetric aperture for the pickup radiation, due to vacuum-chamber misalignments, with a corresponding reduction in the total cooling force as well as a modified transverse profile of the focused UR. The second was increased scattering due to residual gas in the IOTA ring. This would increase the equilibrium transverse beam sizes and single-particle amplitudes, exacerbating the transverse effects described in Section~\ref{sec:simu_transEffects}. Others included nonlinearities in the electron bypass and distorted trajectories in the undulators due to saturation in the steel poles.

 Three-dimensional scans of the integrated experimental apparatus and correspondingly modified simulations of the UR suggest a reduction of the angular acceptance of the lens by $\sim 25\%$ with a corresponding reduction of the maximum kick strength by $\sim 15\%$.  The experiment also observed a weaker-than-expected coupling ratio for the nominal excitation of the QX1 quadrupole magnet, which suggests the presence of additional sources of coupling in the bypass. Reducing the strength of QX1 by half results in the observed experimental coupling ratio of 1.00:0.35. With this reduced coupling, the effects of OSC are seen most prominently in the longitudinal plane. To account for the increased residual gas scattering, we insert a scattering element in the lattice that simulates coulomb scattering events. The equilibrium transverse emittance in the experiment due to SR damping alone was measured to be approximately four times larger than the design value. The scattering strength was tuned to match this empirical value and the same simulations as before are run with these modifications. The results are outlined in Table~\ref{tab:exp_rates}.

\begin{table}[hhh!!!]
    \caption{Damping rates simulated with {\sc elegant} for the IOTA experimental (``as built") configuration.\label{tab:exp_rates}}
    \vspace{3mm}
    \begin{ruledtabular}
    \begin{tabular}{lddd}
      &  \multicolumn{3}{c}{damping rates ($s^{-1}$)} \\
    \textrm{Mode} & \textrm{$\tau_x^{-1}$} & \textrm{$\tau_y^{-1}$} & \textrm{$\tau_s^{-1}$} \\
    \hline
        SR only & 0.9 & 0.95 & 2.03 \\
        \hline
        uncoupled (1D)& 0.9 & 0.95 & 20.75 \\
        s/x coupled (2D)& 3.49 & 0.95 & 17.65 \\
        s/x/y coupled (3D)& 2.38 & 1.80 & 17.40\\
    \end{tabular}
    \end{ruledtabular}

\end{table}

As expected, the overall rates are significantly reduced and the effect is seen mostly in the longitudinal plane. Subtracting the SR-damping rates, the total OSC damping rates in the 1D, 2D and 3D configurations are $18.7$~s$^{-1}$, $17.2$~s$^{-1}$ and $17.7$~s$^{-1}$, respectively. This is in good agreement with the experimentally measured value of $\tau_{tot}^{-1}=18.4$~s$^{-1}$ and suggests that exacerbated transverse effects were the principal source of reduced OSC force in the experiment.

\subsubsection{Optical Delay and Equilibrium Distributions~\label{sec:optdelayequil}}

The standard configuration of OSC is designed to reduce the deviations in the particles' positions and momenta, producing a beam with a lower 6D emittance. This relies on proper temporal alignment of the beam and UR in the kicker such that a reference particle experiences no net energy change. The optical delay system is responsible for establishing and maintaining the correct optical path length. In the IOTA experiment, the optical delay system is made up of two rotating glass plates which provide fine control over the optical delay~\cite{dick-2021-a}. We implement this in the model using a phase term, $\psi_0$, in Eq.~\ref{eq:tt_coh}. 

The normal operation of OSC has a delay phase of $\psi_0=0$ while a delay phase of $\psi_0=\pi$ establishes a ``heating mode" in which the small-amplitude motion is unstable and particles are driven away from the reference momentum. However, because of the periodic nature of the electromagnetic radiation, the particles are pushed toward higher-order attractors in phase space, resulting in ring-like structures in the different phase-space planes.

The IOTA experiment demonstrated the stability of the OSC bypass and the control over the optical delay by slowly sweeping the delay and observing the response of the beam. The optical delay is initially set such that $\delta \ell \sim +N_u\lambda_r$, where there is no overlap between a particle and its corresponding radiation wave packet. The delay plates are then rotated slowly, at a rate of 0.01~deg$\cdot$s$^{-1}$ ($\sim 0.03 \lambda_r\cdot$s$^{-1}$) to a delay length of $\delta \ell \sim -N_u\lambda_r$. This sweep shows the sequence of heating and cooling modes with the beam distribution reaching equilibrium at each point in the scan~\cite{jarvis-2022-a}. The full envelope of the scan constitutes a measurement of the integrated system bandwidth, which was determined to be $\sim 20$~THz.  

\begin{figure}[h!]
    \centering
    \includegraphics[width=8.6cm]{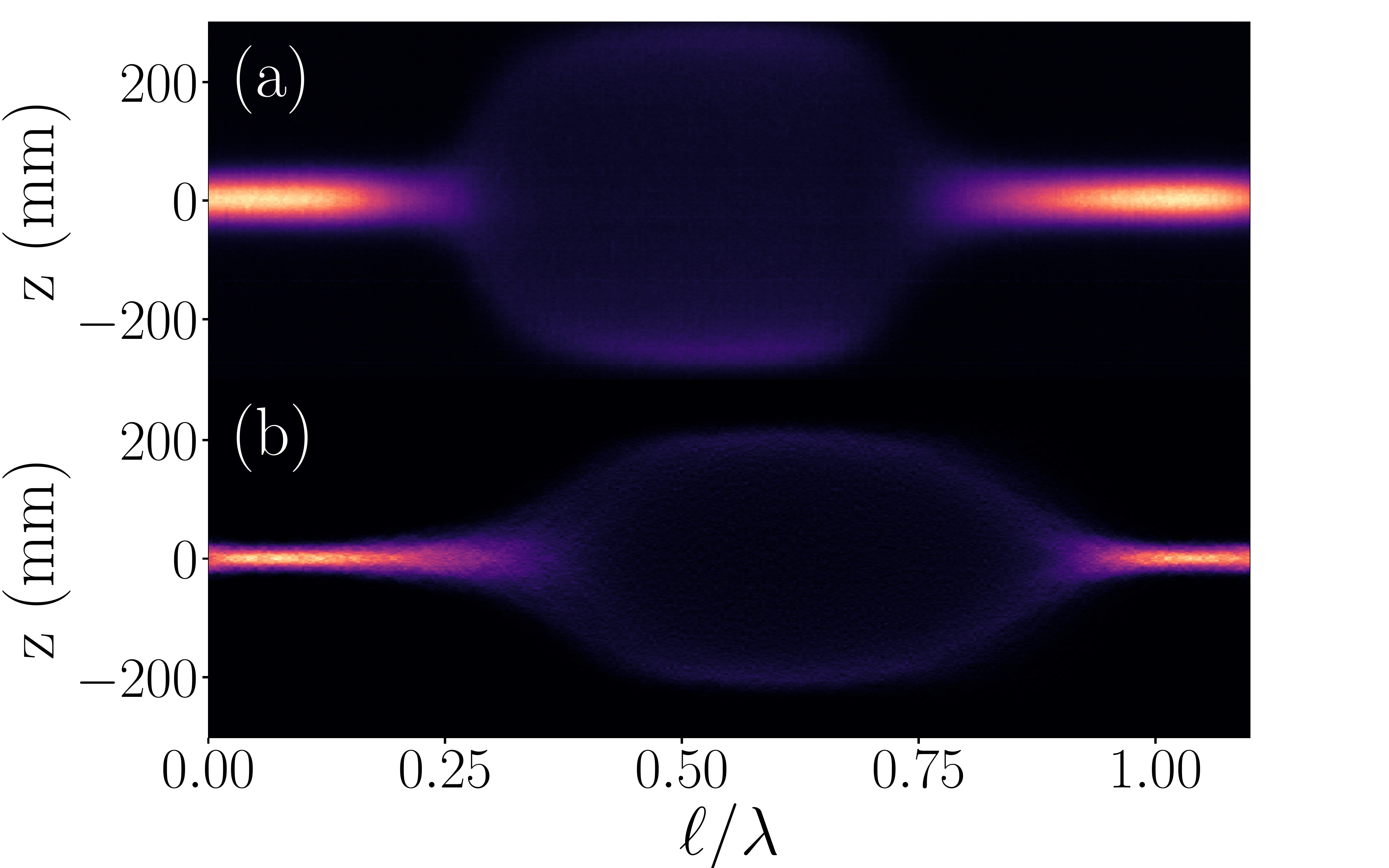}
    \caption{Waterfall plot showing the evolution of longitudinal projections measured (a) and simulated with {\sc elegant} (b) as the optical delay varies within the range  $\delta \ell \in [0,1.25\lambda_r$]. }
    \label{fig:delay_scan}
\end{figure}

In the case of passive OSC, a complete simulation of this scan was not feasible due to the slow scan rate required to ensure that the beam distribution reaches equilibrium for each delay setting~\cite{dick-2022-b}; therefore, we simulate such a scan for a single period $\delta \ell \in [0, \lambda]$. Even so, this shorter scan still required the use of a faster scan rate than the experiment, and some resulting differences are apparent. Figure~\ref{fig:delay_scan} compares the measured and simulated evolution of the longitudinal distribution at equilibrium as the optical delay is varied.
For these simulations, we used 200 macroparticles and executed a slow phase sweep through a single period of the OSC force, starting at $\delta \ell =0$ and ending at $\delta \ell=1.25\lambda_r$, with a scanning speed of  $\simeq 1.39\lambda_r$~s$^{-1}$ (corresponding to a total phase shift of $\Delta \psi_0\simeq 2\pi$ every 600 synchrotron periods). Due to computational restrictions, this sweep was $\sim 600$ times faster than the experiment. As a result, the {\sc elegant} data in Fig.~\ref{fig:delay_scan}(b) shows a lagging response to the optical-delay sweep compared to the measurement displayed in Fig.~\ref{fig:delay_scan}(a) as the beam could not fully equilibrate. However, the simulation qualitatively reproduces the experimental data with cooling occurring at $\sim 0$ and $\delta \ell \simeq \lambda_r$ and maximum heating at $\delta \ell \simeq 0.5 \lambda_r$, albeit with a shift of $\sim 0.25 \lambda_r$ due to the rapid scan rate.  

\begin{figure}[h!]
    \centering
    \includegraphics[width=\columnwidth]{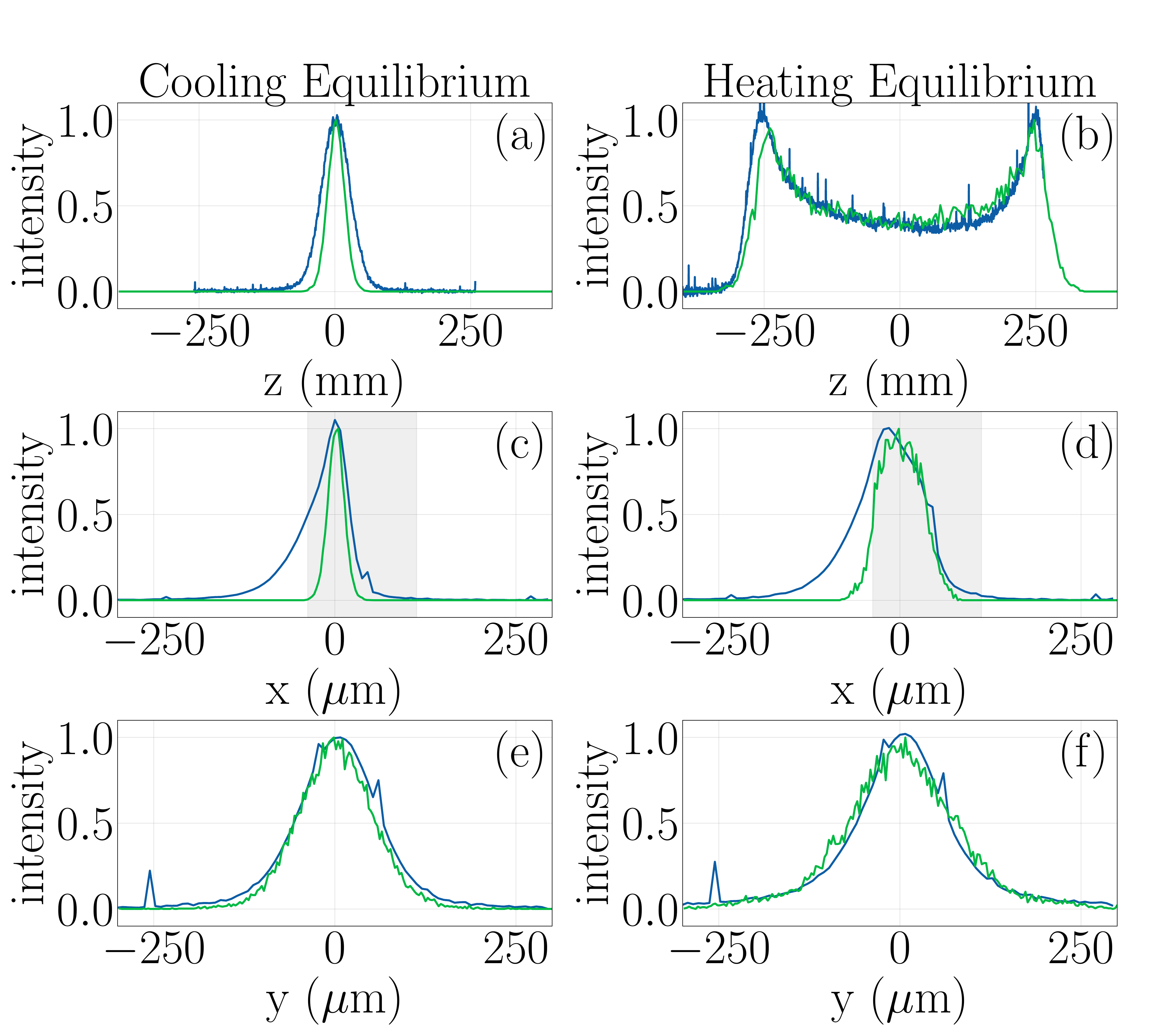}
    \caption{Comparison of the experimental (blue) and simulated (green) equilibrium distributions in longitudinal (a,b), horizontal (c,d) and vertical (e,f) directions for the cooling ($\psi_0=0$, left column) and heating ($\psi_0=\pi$, right column) modes. The shaded area in plots (c,d) represents the regions where the measurements are accurate (see text for details). The development of a tail toward negative values of $x$ is a measurement artifact; see text for details.}
    \label{fig:equilibrium_dist}
\end{figure}

To further investigate the delay variable we simulated equilibrium beam distributions for delay settings of $\psi_0=0$ and $\psi_0=\pi$, which correspond to the OSC cooling and heating modes respectively. The results are compared with the equilibrium distributions recorded in the experiment and are shown in Fig.~\ref{fig:equilibrium_dist}. The transverse distributions were measured in {\sc elegant} by recording particle coordinates in the center of the M2R bending magnet and binning over several thousand turns. The horizontal distributions [see Figs.~\ref{fig:equilibrium_dist}(c) and ~\ref{fig:equilibrium_dist}(d)] in the experiment are asymmetric due to depth-of-field effects and the curvature of the beam trajectory in the main dipoles. The gray shading indicates regions where these effects are reduced and the simulated and experimental distributions can be directly compared. Diffraction effects are present in the experimental data for both transverse planes but are only significant ($\sim 25$\%) for the horizontal plane in the OSC cooling mode. The transverse profiles show good agreement in both cases. 

The longitudinal distributions were recorded at the M3R dipole and compared with streak camera projections from the experiment. The longitudinal heating equilibrium distributions show very good agreement indicating that the computational model behaves as expected when operating in a heating mode. The simulated longitudinal distribution for the cooling case is slightly more narrow than the experiment. This is most likely due to the increased IBS in the cooling mode, as seen in the previous section.

\subsection{Heating Mode Dynamics} 

In the nominal 2-D cooling configuration of OSC (shared cooling between the longitudinal and horizontal planes), the particles are cooled simultaneously in both planes towards the design orbit. However, as discussed in Section~\ref{sec:optdelayequil}, when the optical delay is set to $\psi_0 = \pm \pi$, particles at low amplitudes are driven away from the design orbit towards stable, high-amplitude orbits in phase space. 

In order to quantitatively describe this effect, it is useful to rewrite Eq.~\ref{eq:tt_coh} to also include betatron oscillations. Following Ref.~\cite{lebedev-2014-c}, Eq.~\ref{eq:tt_coh} is parameterized in terms of normalized betatron and synchrotron amplitudes and phases, $(a_x,\Psi_x)$ and $(a_p,\Psi_p)$ respectively, as 
\begin{eqnarray}
        \Delta {\cal E }_i = - {\cal K} \sin[a_x \sin(\Psi_x + \Phi_c) + a_p\sin(\Psi_p) + \psi_0], \nonumber
\end{eqnarray}
where $\Phi_c$ is the phase shift between the momentum kick and betatron motion. Using this definition, integrating over the synchrotron and betatron oscillations reveals the presence of stable-orbit points (or attractors) in the $(a_p,a_x)$-parameter space~\cite{lebedev-2014-a}. The synchrotron amplitude is 

\begin{equation}
    a_p = k \left(M_{51}D + M_{52}D' + M_{56}\right)\delta_m,  \nonumber
\end{equation}
where $\delta_m$ is the maximum fractional-momentum deviation over a full synchrotron period. The normalized betatron amplitude is
\begin{eqnarray}
    a_x = k \sqrt{\tilde\epsilon_x\left(\beta_x M_{51}^2 - 2\alpha_x M_{51}M_{52} + (1+\alpha_x^2)M_{52}^2/\beta_x\right)}, 
    \nonumber
\end{eqnarray}
where $\tilde\epsilon \equiv \frac{1}{\beta_x}[x^2+(\alpha_x x + \beta_x x')^2]$ is the Courant-Snyder invariant, and $(\beta_x, \alpha_x)$ are the usual Courant-Snyder parameters~\cite{lee-2018-a}. 
\begin{figure}[hhhhhhhh!!!!!!!!]
    \centering
    \includegraphics[width=\columnwidth]{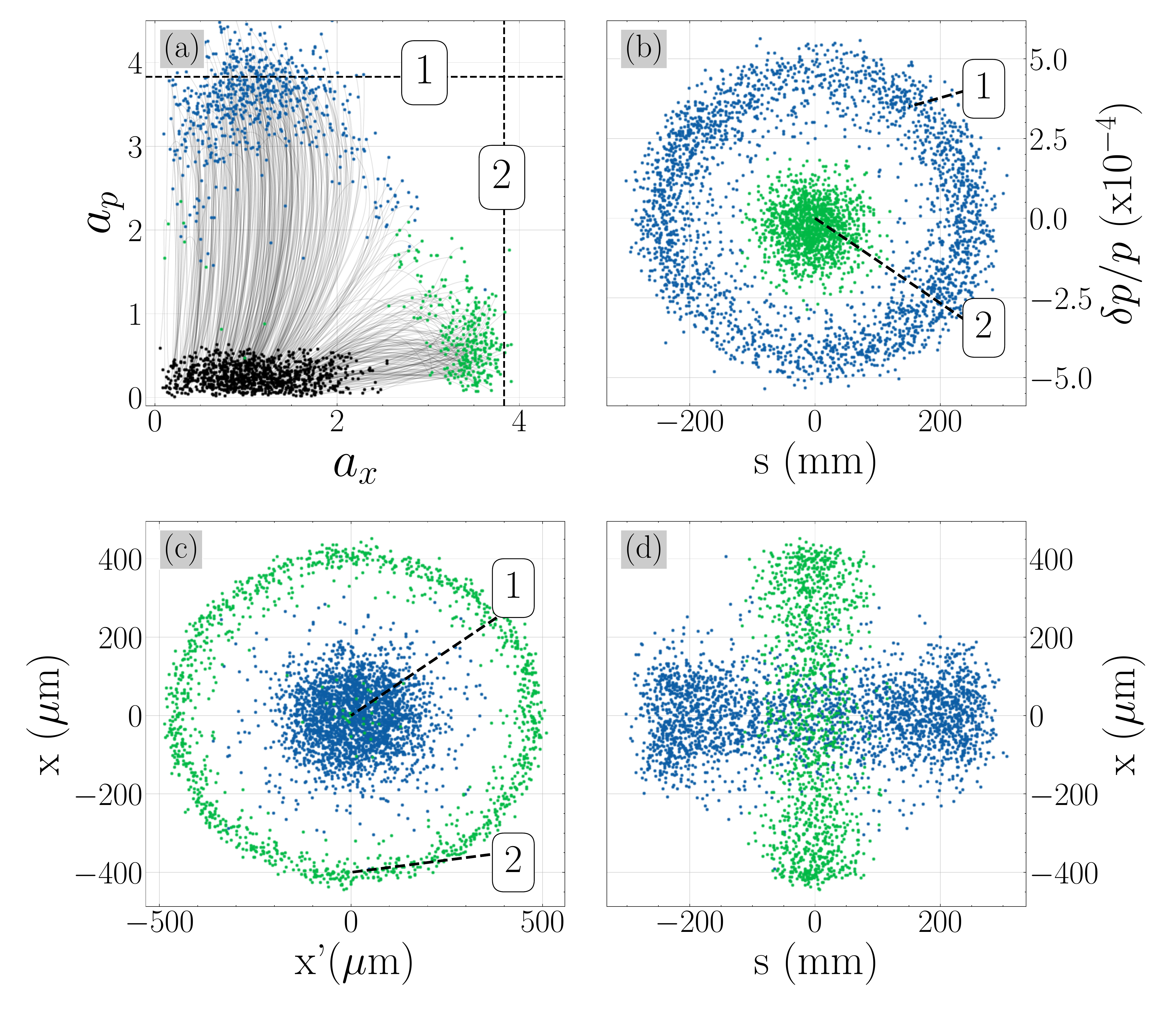}
    \caption{Evolution of macroparticles from their initial positions (black dots) in ($a_p$-$a_x$) amplitude space due to OSC heating in the 2D-cooling configuration (a). The blue dots labeled ``1" represent macroparticles associated with the high longitudinal-amplitude attractor, while the green dots labeled ``2" correspond to macroparticles in the high horizontal-amplitude attractor. The equilibrium distribution is shown in longitudinal (b) and horizontal phase space (c). The spatiotemporal distribution $(s,x)$ of particles is shown in (d), where the axes correspond to the spatial components of (b) and (c). The attractors ``1" and ``2" respectively correspond to ($a_x=0, a_p\simeq 3.81$) and  ($a_x\simeq 3.81, a_p=0$) where 3.81 corresponds to the first zero of the $J_1$ Bessel's function. }
    \label{fig:long_heatingdist_phaseplot}
\end{figure}

In the heating mode ($\psi_0=\pm \pi$) for the 2-D configuration, the lowest-order attractors have high amplitude in one phase plane and simultaneously low amplitude in the other. This feature was observed in the {\sc elegant} simulations by examining the evolution of the horizontal and longitudinal amplitudes associated with randomly sampled macroparticles within the bunch; see Fig.~\ref{fig:long_heatingdist_phaseplot}(a). The macroparticles evolve to one of the two heating attractors with either high-amplitude (blue dots) or low-amplitude (green dots) longitudinal motion. The bifurcation of the ensemble towards these attractors results in an equilibrium distribution with a "bullseye"-shaped distribution in the longitudinal and horizontal phase-space consisting of a bright core with a peripheral ring as shown respectively in  Fig.~\ref{fig:long_heatingdist_phaseplot}(b) and (c). These bifurcated phase space distributions yield a multi-modal spatiotemporal $(s,x)$ distribution as shown in Fig.~\ref{fig:long_heatingdist_phaseplot}(d). The ratio of macroparticles populating these regions of the $(a_p,a_x)$ parameter space depends on the strength of the longitudinal-to-horizontal coupling (1:0.34 in this case). For an imbalanced ratio, the particles will be preferentially driven to the stronger attractor, which results in cooling for the other phase plane.  This effect was clearly observed in the OSC experiment \cite{jarvis-2022-a}.

\begin{figure}[h!]
    \centering\includegraphics[width=\columnwidth]{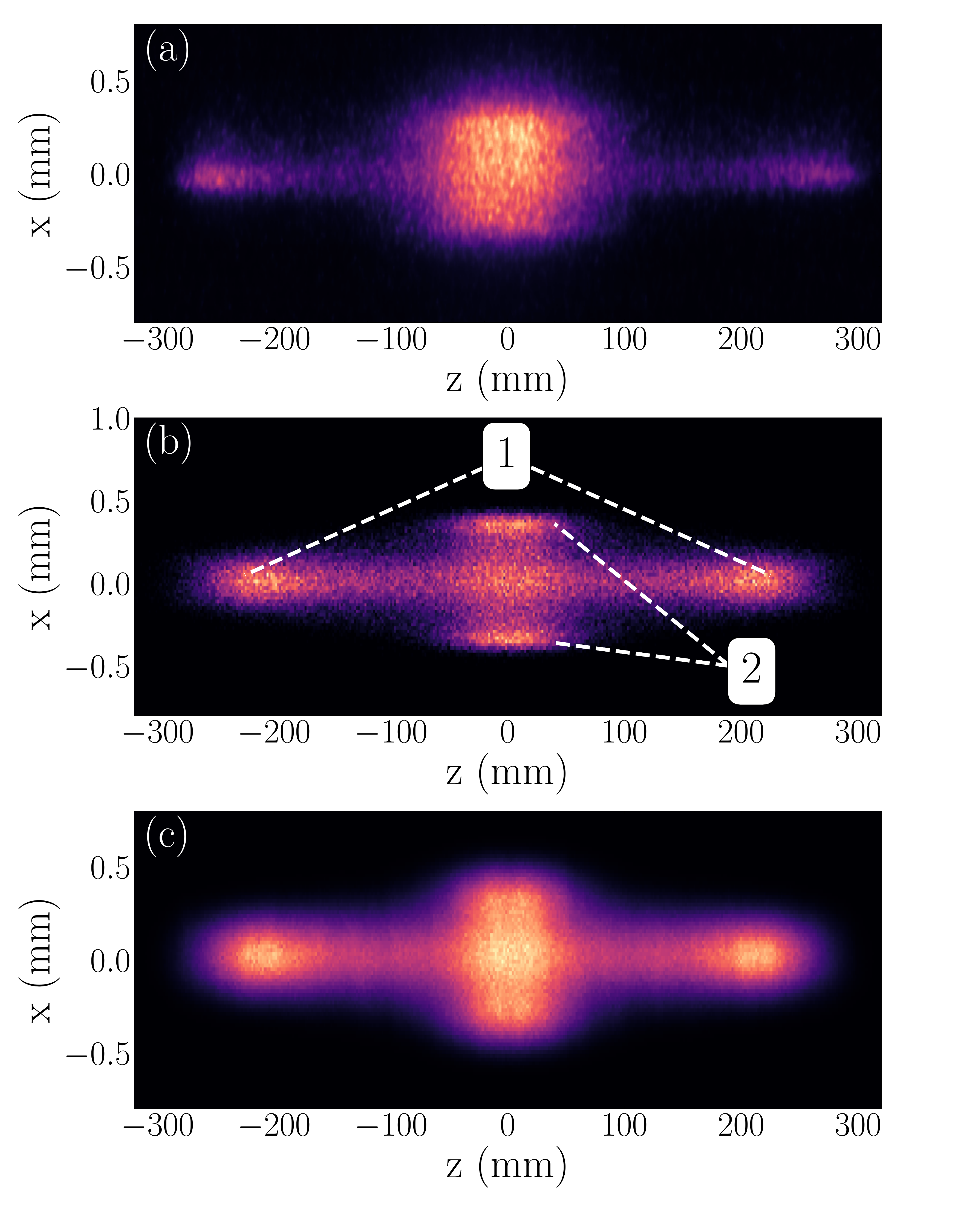}
    \caption{Streak camera measurement of the spatiotemporal distribution $(s,x)$ at M3R for the 2D-cooling configuration ($s-x$ coupling) operated in the heating mode (a) and corresponding distribution simulated with {\sc elegant}  (b) along with the simulated measurement obtained by smearing the simulated distribution with a Gaussian point-spread function along the $x$ axis. The labels ``1" and ``2" in (b) refer to the same attractors as defined in  Fig.~\ref{fig:long_heatingdist_phaseplot}.}
    \label{fig:streak_dists}
\end{figure}

In the experiment, the equilibrium spatiotemporal distribution was recorded at M3R for the 2-D cooling configuration and is shown in Fig.~\ref{fig:streak_dists}(a). {\sc Elegant} simulations were performed for the same configuration and the macroparticle coordinates were recorded at the effective source plane of the M3R diagnostics.

The results of this simulation appear in Fig.~\ref{fig:streak_dists}(b,c). Figure~\ref{fig:streak_dists}(b) shows the bifurcation of the beam as particles move to one of the two high-amplitude attractors as already examined in Fig.~\ref{fig:long_heatingdist_phaseplot}(d). The simulated streak camera image confirms the presence of two lobes along the longitudinal axis corresponding to the high-synchrotron-amplitude, low-betatron-amplitude attractor ($a_x=0, a_p\simeq \nu_{11}$) where $\nu_{11}\simeq 3.81$ represents the first zero of the Bessel function of the first kind $J_1(x)$~\cite{lebedev-2014-a}. The two lobes along the ordinate of Fig.~\ref{fig:streak_dists}(b) correspond to the other attractor ($a_x\simeq \nu_{11}, a_p=0$). The fine features observed in the simulated distribution shown in Fig.~\ref{fig:streak_dists}(b) are not present in the experimental data [Fig.~\ref{fig:streak_dists}(a)] owing to the finite resolution and other limitations of the streak camera imaging system. To make a qualitative comparison, a Gaussian point-spread function was applied to the simulated distribution along the $x$ dimension. This simulated ``measurement" appears in Fig.~\ref{fig:streak_dists}(c) and agrees well with the experimentally-measured distribution of Fig.~\ref{fig:streak_dists}(a) thus confirming the ability of the numerical model to accurately simulate the dynamics of heating modes as the beam evolves toward equilibrium. Where direct comparisons are possible, we find the simulated beam distributions to be in good quantitative agreement with the measured distributions.

\section{Additional Features}

The OSC model was developed to be a general tool, and it includes several features that could not be easily studied in the IOTA experiment. The initial experiment was designed to use beams with low bunch charge to reduce the effects of IBS. This also has the effect of reducing the incoherent contributions to OSC as very few particles (on average $N_s\in[20,2000]$) will populate each sampling slice. There are other features of the model which can be demonstrated with minimal additional setup such as amplified OSC and errors in angular alignment using multiple OSC PU-KU pairs. In this section, we present brief examples and analysis of these features.

\subsection{Incoherent OSC}
\begin{figure}[t]
    \centering
    \includegraphics[width=1\columnwidth]{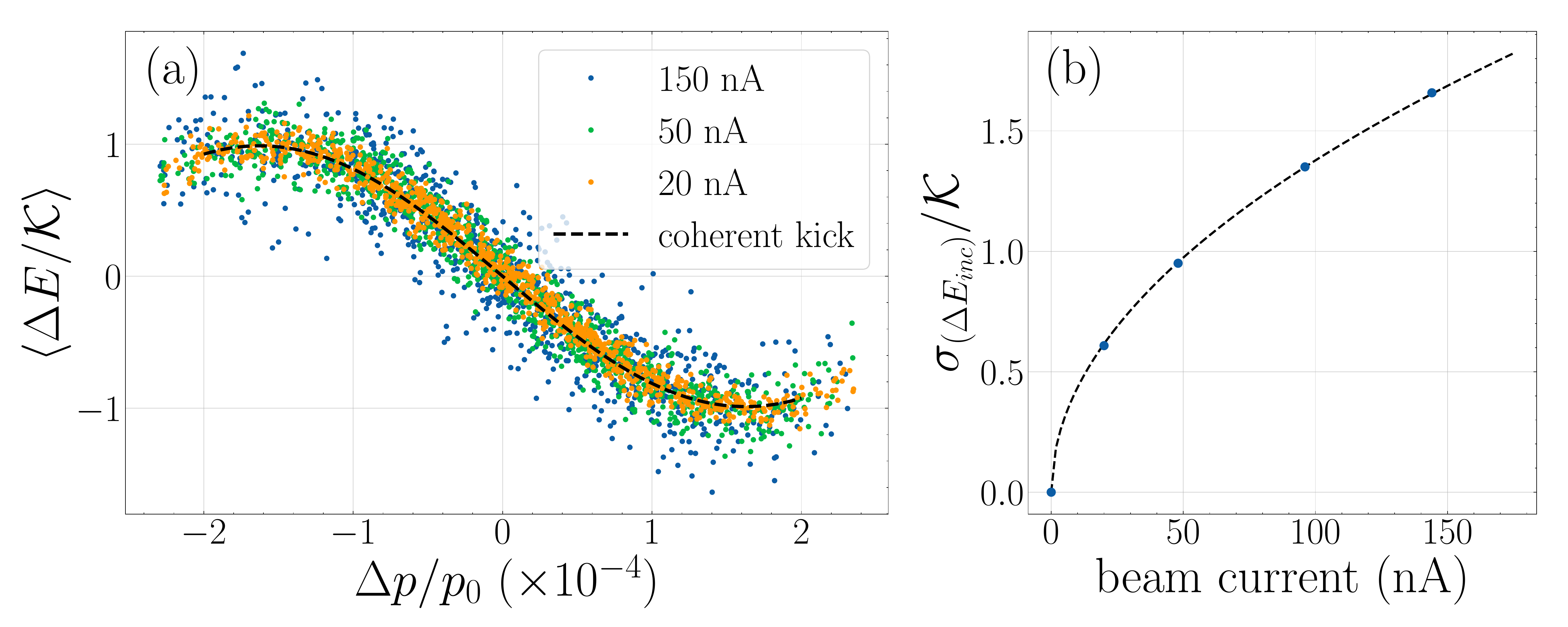}
    \caption{The total energy kick $\Delta E \equiv \Delta {\cal E} +\Delta \tilde{\cal E}$, including the coherent and incoherent contributions, each particle receives relative to the maximum kick ${\cal K}$ (a) and the standard deviation of the incoherent contribution for a single turn normalized to the maximum kick strength(b).}
    \label{fig:incoherent}
\end{figure}

The incoherent contributions of neighboring particles have been modeled before using a statistical approach \cite{wang-2021-a} but have not been experimentally studied. Our model computes this effect directly by applying a kick to the individual particle from every particle in its sample slice. Figure~\ref{fig:incoherent}(a) shows the average kick each particle in a beam receives as a function of its average momentum deviation. The values are averaged over 50 turns, which corresponds to elapsed time of $0.67$~ms (i.e. $0.3\%$ of a synchrotron period). The change in longitudinal position and momentum of a single particle is negligible over this duration, which provides a reasonable average kick for a given momentum deviation. We simulated this effect for three different beam currents. As the charge density increase, the incoherent effects become more dominant. Figure~\ref{fig:incoherent}(b) shows the standard deviation of the incoherent kick that a particle will experience in a single turn. The variance $\sigma_{\text{inc}}^2$ grows linearly with the longitudinal charge density. The average standard deviation scales with the number of turns $N$ as $\langle\sigma_{\text{inc}}\rangle_N=\sigma_{\text{inc}}/\sqrt{N}$ so this effect is negligible for low charge and long bunch length.

\subsection{Amplified OSC}
\begin{figure}[h!]
    \centering
    \includegraphics[width=1\columnwidth]{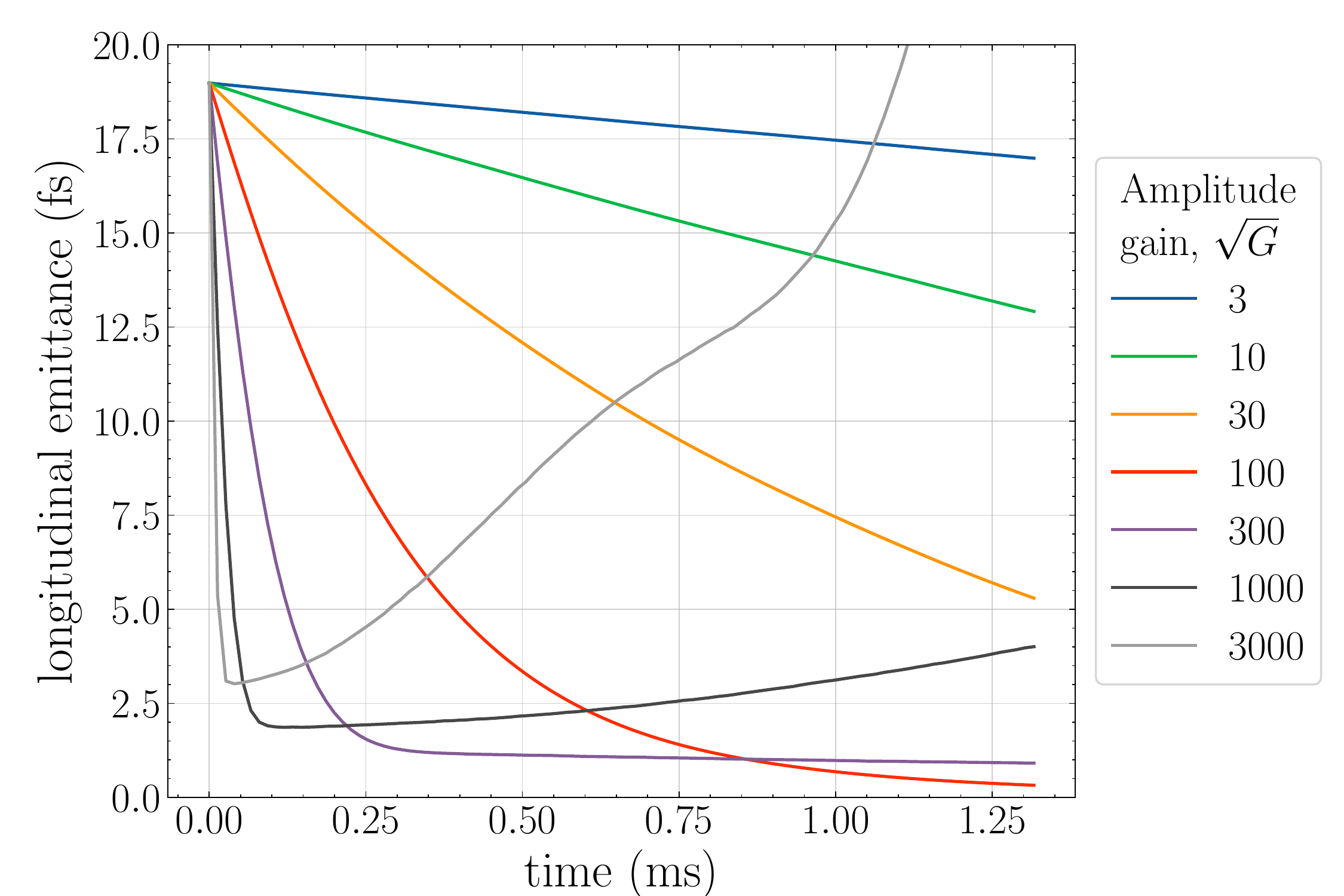}
    \caption{Effect of amplification on OSC damping for gains $G$ of 10-70\,dB.  The simulation used an amplitude gain, $\sqrt{G}$.}
    \label{fig:amplified}
\end{figure}

In future experiments, the UR produced in the pickup will be amplified before interacting with the particles in the kicker. By amplifying the PU radiation, the beam can be cooled much faster than in the passive case.  In the model, amplification is included by simply increasing the kick strength parameter.  The limited optical delay of the initial IOTA experiment could not accommodate an optical amplifier; however, for purposes of demonstration, Fig.~\ref{fig:amplified} presents the simulated evolution of the longitudinal emittance $\varepsilon_t \equiv \varepsilon_s/c$ as the beam is cooled by OSC; see Eq.~\ref{eq:emit}. These calculations used the nominal uncoupled IOTA lattice and consider various optical-power gains $G$ ranging from 10 to 70\,dB (corresponding to $\sqrt{G} \in[3, 3000])$. The simulations include incoherent effects but neglect IBS.  
When incoherent effects are negligible, the cooling rate increases linearly with the maximum energy kick (proportional to $\sqrt{G}$). In the absence of size-dependent diffusion effects like IBS, the longitudinal emittance in equilibrium follows the relationship  $\varepsilon_{t,OSC} / \varepsilon_{t,SR} = \tau_{OSC}/\tau_{SR}$~\cite{jarvis-2022-a}.  As the gain increases, the incoherent contributions limit the longitudinal emittance. Eventually, large gains yield an unstable particle distribution  with increasing longitudinal emittance.  The simulations indicate that an optimal gain for providing the lowest-emittance beam while maintaining equilibrium in the IOTA lattice is $G \simeq 40$~dB.

\subsection{Multiple pairs of OSC elements}
The OSC model discussed so far consists of a pair of lattice elements, the {\tt pickup} and the {\tt kicker} linked using an ID-string within {\sc elegant}. Specifically, the results discussed in Section~\ref{sec:benchmarking} use a single OSC pair considered as thin elements. However, multiple pairs can be defined so as to introduce an OSC element for each undulator period (i.e. in a sequence of paired {\tt pickup} and {\tt kicker} elements). Such a configuration allows for the investigation of thick-lens effects in the OSC elements. 

One such effect is angular misalignment in the PU and KU. Similar to the discussion in Section~\ref{sec:singleparticleTransEffect}, a beam entering the kicker with at an angle will experience the off-axis electric field in the KU. This generally will reduce the cooling damping decrement. 
\begin{figure}[hhhhhh!]
    \centering
    \includegraphics[width=1\columnwidth]{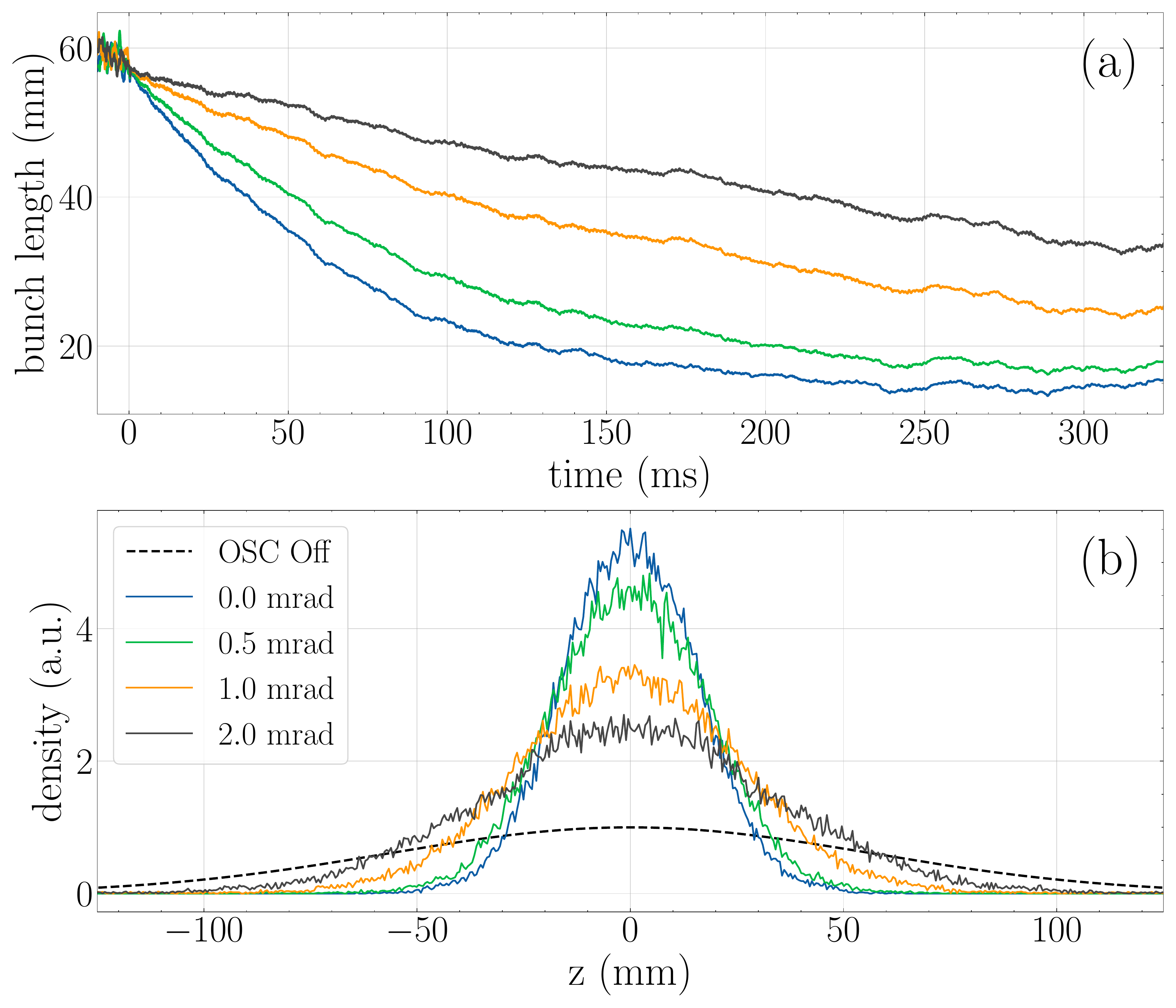}
    \caption{Effects of angular misalignment on damping (a) and equilibrium longitudinal distribution (b).}
    \label{fig:angular}
\end{figure}
Figure~\ref{fig:angular} presents the evolution of the bunch length of a beam in the presence of OSC with increasing misalignment in the KU. The angular misalignment $\theta$ is introduced by assigning a transverse misalignment $\delta x_n=(s_n-s_{KU})\tan\theta$ to each {\tt kicker} element (indexed by the $n$ subscript) depending on its location within the KU. Angular misalignment decreases the cooling rate and results in a longer equilibrium bunch length. In the example shown in Fig.~\ref{fig:angular}, corresponding to the IOTA passive-OSC case, a $\sim 2$~fold increase in the equilibrium bunch length and a $\sim30\%$ reduction in cooling rate is observed for an angular misalignment $\theta =500$~\textmu{}rad.

\section{Conclusion}

We have developed a fast computational model of the OSC mechanism and have demonstrated its ability to accurately simulate the dynamics of the passive-OSC proof-of-principle experiment performed at the IOTA storage ring. 
We modeled the IOTA storage ring in {\sc elegant} and observed good quantitative and qualitative agreement between the simulations and the measured performance of the OSC system for a variety of system configurations. The current model includes the effects of the transverse profile of UR, optical delay, and incoherent contributions to OSC of neighboring particles. Other features of the model (incoherent kicks, effects of misalignment, and optical gain) were also showcased and will guide future OSC experiments while providing a microscopic understanding of OSC dynamics.  

Finally, although the current model implemented in {\sc elegant} considers a transit-time OSC configuration, it could be easily extended to implement other stochastic cooling configurations and, more broadly, be used to investigate the control of beam distributions using self-field radiated at an earlier time.

\section{Acknowledgements}
This work was supported by U.S. National Science Foundation under the award PHY-1549132, the Center for Bright Beams. Fermilab is managed by the Fermi Research Alliance, LLC for the DOE under contract number DE-AC02-07CH11359. The work of A.J.D. was partially supported by the DOE Office of Science Graduate Student Research (SCGSR) Program. The computing resources used for this research were provided on {\sc bebop},  a high-performance computing cluster operated by the Laboratory Computing Resource Center (LCRC) at Argonne National Laboratory.

\bibliography{oscBenchmarking}

\end{document}